# Role of the annealing parameters on the resistance of indium tin oxide nanocrystalline films

Fabio Marangi[1,2]*, Luigi Fenzi[1,2]*, Michele Bellingeri[3,4], Francesco Scotognella[1,2]#

[1] Department of Physics, Politecnico di Milano, 20133 Milan, Italy

[2] Center for Nano Science and Technology@PoliMi, Istituto Italiano di Tecnologia (IIT), 20133 Milan, Italy

[3] Dipartimento di Scienze Matematiche, Fisiche e Informatiche, Università di Parma, via G.P. Usberti, 7/a, 43124 Parma, Italy

[4] INFN, Gruppo Collegato di Parma, I-43124 Parma, Italia

* The authors equally contributed to this work.

# E-mail address of the corresponding author: francesco.scotognella@polimi.it

## Abstract

The optical and electrical properties of films made of nanoparticles of indium tin oxide (ITO) are widely studied because of the significance of this material for transparent electrodes, smart windows, and nonlinear optics components. In this work, a systematic study of the resistance in ITO nanocrystalline films, as a function of post-fabrication parameters, such as the temperature and time of annealing, has been performed. A tunability of the resistance with the annealing parameters, in a range of three orders of magnitude, has been demonstrated.

## Introduction

Transparent conductive oxides (TCO) are interesting materials since they show good conductivity and they absorb in the near infrared region. For this reason, they are vastly employed for the fabrication of electrodes in organic solar cells, quantum dot solar cells, metal halide perovskite solar cells, and so on [1–3]. Moreover, because of their strong absorption in the near infrared, transparent conductive oxides are being studied as potential materials for infrared photovoltaics [4]. Since their absorption is not due to interband transition, but to plasma frequency, the working principle of a TCO-based solar cell is not the traditional photovoltaic effect: Electromagnetic radiation, that impinges the material, generates hot electrons; at the interface with a proper semiconductor, hot electron extraction from the TCO to the semiconductor can occur if the energy of the hot electron is higher with respect to the bottom of the conduction band of the semiconductor [5]. This phenomenon has been observed in indium tin oxide (ITO)/tin oxide interfaces [6], fluorine indium co-doped cadmium oxide/rhodamine interfaces [7], and indium tin oxide/molybdenum disulphide interfaces [8]. Finally, exploiting the possibility to tune the plasmon resonance, via a change of the number of carriers in the material, TCOs are widely employed as smart windows [9–12].

TCOs films are usually fabricated via sputtering or chemical vapor deposition. However, the large-scale fabrication of stable TCO nanoparticles [13,14] in dispersion allows low-cost deposition techniques, such as spin coating [15,16]. The optical properties of ITO nanocrystalline films have been carefully studied by Mendelsberg et al. [17] and Sygletou et al. [18].

Here, the electrical properties of ITO nanocrystalline films have been studied. The films have been fabricated via spin coating starting from nanocrystal colloidal dispersions. The resistance has been carefully studied as a function of the temperature and time of annealing. With the control of the two annealing parameters, the resistance has been tuned in a range larger than three orders of magnitude.

## Methods

**Fabrication procedure**: Fabrication of the samples was performed starting from a diluted dispersion of commercial ITO nanocrystals (NCs) (GetNanoMaterials ITO-200-30WT, indium tin oxide ITO nanoparticles 99.99%, 20 – 30 nm, $In_2O_3 : SnO_2$ 90:10 wt%, 30 wt% in water). The NCs were deposited on Oxygen plasma activated glass substrates using spin coating (1500 RPM, 60s). Samples were annealed in the open air on a hotplate to get rid of organic ligands and provide heat treatment to the films to favor reorganization and obtain a better packing of the NCs. Heat treatment temperatures ranged from 150 °C to 400 °C. In order to have reliable data, a number of three samples for each annealing temperature have been fabricated. The time of annealing varied from 30 min to 3 h. The thickness of the film is between 50 and 70 nm [18].

**Light absorption measurements**: The absorption spectra of the ITO NCs films have been measured with a Perkin Elmer Lambda 1050 spectrophotometer equipped with deuterium (280–320 nm) and tungsten (320–3300 nm) lamps. The signal is collected by three detectors acting in different spectral regions (photomultiplier [180, 860] nm, InGaAs [860, 1300] nm and PbS [1300, 3300] nm).

**Electrical characterization**: The electrical conductivity of the ITO nanocrystalline films has been measured via the four-point probe technique [19]. The square configuration of the probes was employed, due to its higher sensitivity and the smaller measured area of the sample. In order to achieve accurate results, micro-pads of silver were deposited at 60 °C using inkjet-printing on top of the ITO layer (all at the same distance of 1 mm), providing not only better electrical contact between the probes and the samples, but also the fixed probe spacing to be inserted in the resistance formula:

$$R = R_{ITO} = R_{ITO}^{square} = \frac{2\pi}{\ln 2}\frac{V_2 - V_3}{I_1} \qquad (1)$$

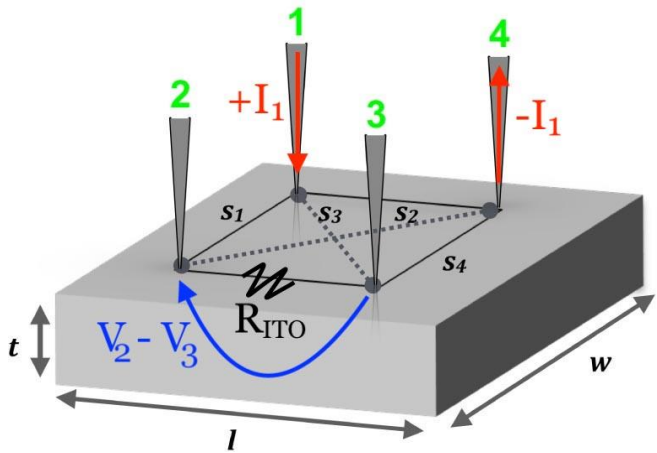

Figure 1. Sketch of the ITO nanocrystalline film and the four-point probe technique to measure the resistance.

Once the probes were positioned onto the silver micro-pads, a current was sent through probe 1 inside the sample. In Figure 1 is possible to notice the path followed by the current inside the sample. In the four-point probe method, the resistance becomes a function only of the impinging current and of the floating potential of the inner pair ($V_2 - V_3$), therefore the contribution to the sample resistance provided by the probes themselves can be neglected [19]. As a matter of fact, the impinging current intensity (and therefore the applied reliance potential) firstly is increased, starting from a certain chosen value until it reaches the opposite one ($\pm 10$ nA – 1 $\mu$A). Once the maximum value of current $I_1$ is achieved, the current starts decreasing, retracing exactly its footsteps. Therefore, the sample response is measured in terms of the opposing resistance to the current flow (i.e., $R$).

## Results and Discussion

In Figure 2, the absorption spectra of the ITO nanocrystalline films annealed at 150 °C for different annealing times have been shown. From the sample without annealing ("NoAnn" in Figure 2) to the sample annealed for 180 minutes (3h, dark green curve in Figure 2) a red shift of the plasmonic resonance of the ITO film is observed. The shifts in the absorption peak might be correlated to oxidation phenomena occurring during the heating treatment, which hinder the capability of the charges to move freely in the material. Oxygen fills the structural vacancies in the ITO film, entrapping the free carriers and therefore reducing the localized surface plasmon resonance (LSPR) phenomenon [20].

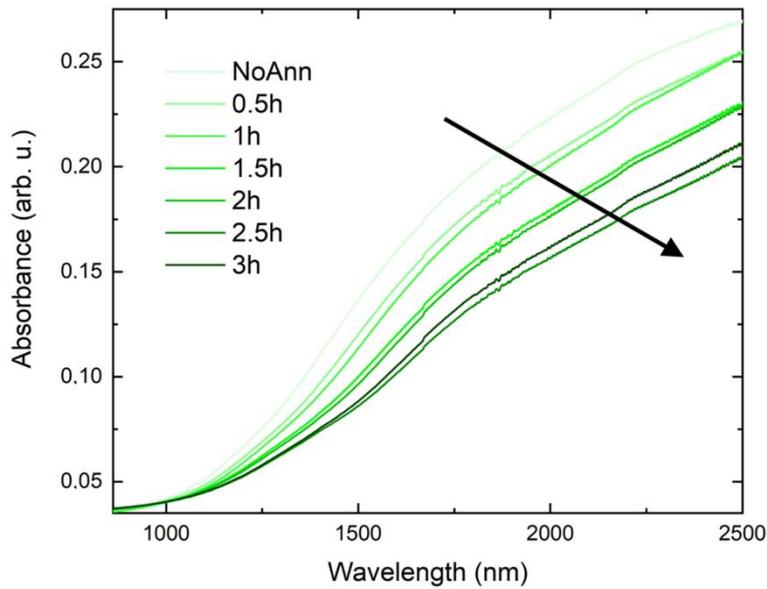

Figure 2. Absorption spectra of the ITO nanocrystalline film annealed at 150 °C as a function of the annealing time. The arrow indicates the increase in annealing time.

Figure 3 reports the absorption spectra of the ITO nanocrystalline films annealed for 30 minutes at different temperatures. Increasing the temperature of annealing has a stronger influence on the plasmonic resonance. In fact, the red shift of the absorption band associated with the plasmon is greatly enhanced. At a given wavelength (i.e., 2000 nm) the absorption could be as much as 80% less. The control of the combination of the two parameters (temperature and time) allows extreme tuning of the optical properties of the films. As will be later discussed, those parameters also influence resistivity of the films. The temperature that corresponds to the lowest absorption intensity at 2500 nm is 350 °C. Moreover, the influence of the temperature on the absorption is stronger with respect to the influence of the annealing time.

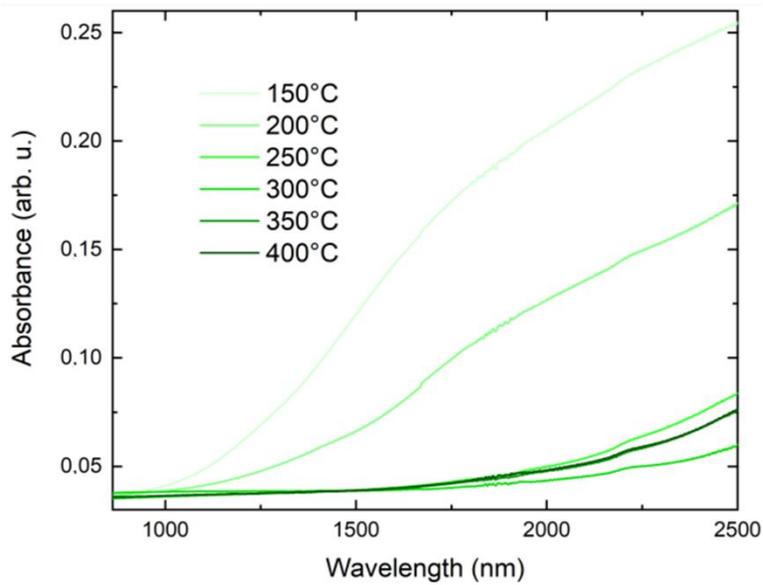

Figure 3. Absorption spectra of the ITO nanocrystalline films, annealed for 30 minutes, as a function of the annealing temperature.

In Figure 4, the resistance values of the ITO nanocrystalline films as a function of the annealing temperature and the annealing time have been plotted. In Figure 5, we depict the average resistance values as a function of the sole annealing temperature. The annealing time has a marginal influence to determine the resistance of the ITO nanocrystalline films, and the resistance shows a slight decrease as a function of the annealing time. On the contrary, the resistance of the ITO nanocrystalline films strongly decreases as function of the temperature of annealing.

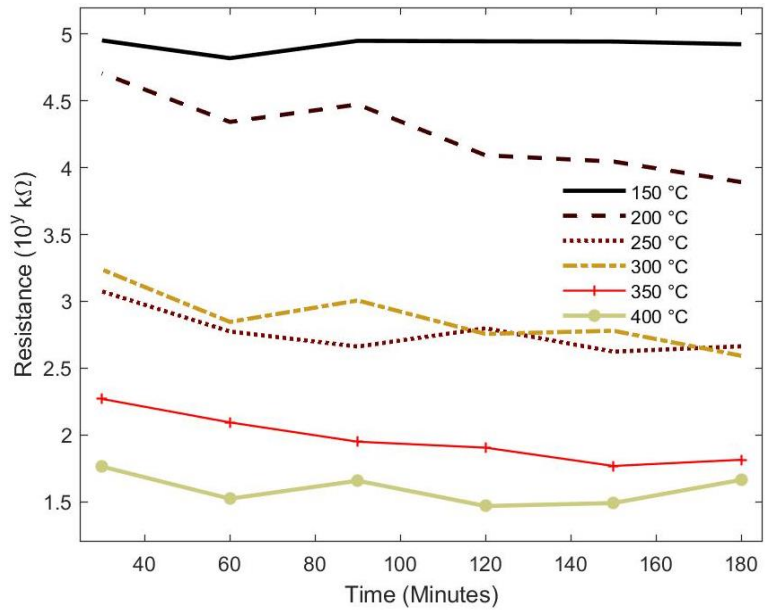

Figure 4. Resistance of ITO nanocrystalline films (in logarithmic scale, e.g. 2 corresponds to $10^2$ kΩ), for different annealing temperatures (from 150 °C to 400 °C), as a function of the annealing time (from 30 minutes to 180 minutes). Each point is the average of 2 or 3 samples.

It is noteworthy that the resistance values for the ITO films annealed at 250 °C and 300 °C are very similar. We could ascribe this phenomenon to a phase transition in which the absorption of heat is related to the phase change of latent heat. The range of temperature is in fair agreement with the phase transition temperature reported by Song et al [21].

We would like to highlight that, with an ITO nanocrystalline film grown with the nanoparticle dispersion and with the same fabrication conditions, it is possible to modify the resistance of the film from a maximum of 89.77 MΩ to a minimum of 29.33 kΩ, spanning more than three orders of magnitude (Figure 4).

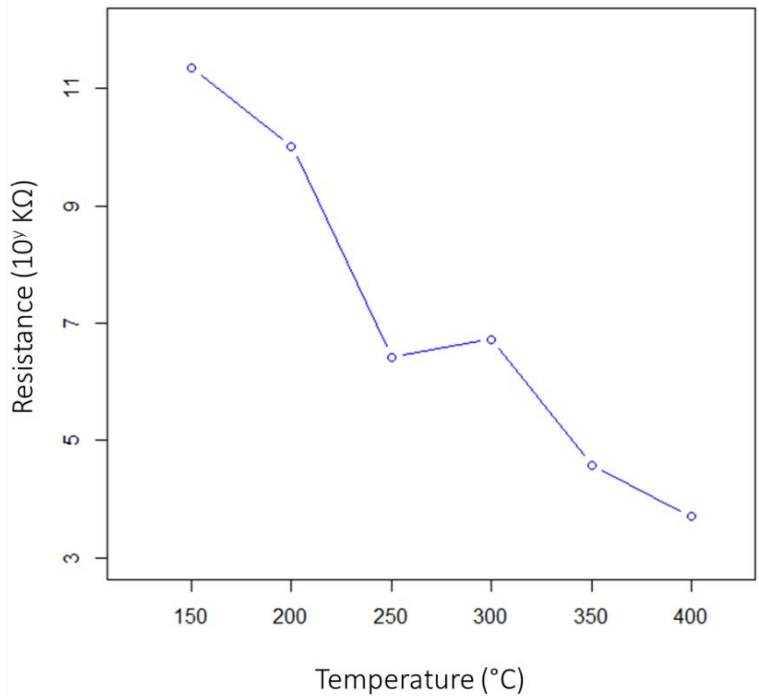

Figure 5. Average resistance of ITO nanocrystalline films along the annealing time (from 30 minutes to 180 minutes) for different annealing temperatures (from 150 °C to 400 °C). The resistance is in logarithmic scale, e.g. 2 corresponds to $10^2$ kΩ. Each point is the average of 2 or 3 samples.

## Conclusion

In this study ITO nanocrystalline films have been fabricated via spin coating, starting from nanocrystal colloidal dispersions. By fabricating samples with the same nanocrystal batch and with the same fabrication conditions, very different resistance values have been achieved, spanning more than three orders of magnitude. The tunability of the resistance with annealing parameters could be very interesting to fabricate devices with desired optical and electrical properties. Knowing the parameters affecting the electrical properties of the films could be of great importance during operation for some applications. For example, excessive heating, when this material is used for screens, photovoltaics, and so on, can alter the optical and electrical properties and affect the functionality of the device. On the other hand, with appropriate fabrication techniques (e.g., laser heating), selective annealing of the nanocrystalline layer can be achieved, favoring more conductive paths and creating patterns for charge transport. Selective annealing of the surface, in bulkier films, could be exploited for desired interface properties. If the electrical properties do not matter, "heat patterning" can be used to alter the optical properties of the film to enhance or reduce absorption. Such a possibility may result in a game-changing for photovoltaics when the integration of one or more devices is involved.

**Acknowledgment**

This project has received funding from the European Research Council (ERC) under the European Union's Horizon 2020 research and innovation programme (grant agreement No. [816313]) and under Horizon Europe (grant agreement No. [101061820]).